\documentclass[aps,prl,twocolumn,floatfix,amssymb,showpacs]{revtex4}
\usepackage{amsmath,pifont}
\usepackage{epsfig}
\newcommand{\pe}[1]{#1_{\bot}}
\newcommand{\pa}[1]{#1_{\|}}
\begin{document}
\title{Critical Behavior of the Two-Dimensional Randomly Driven Lattice Gas}
\author{Sergio Caracciolo$^{1}$, Andrea Gambassi$^{2}$, Massimiliano
Gubinelli$^{3}$, and Andrea Pelissetto$^{4}$}
\affiliation{$^1$Dipartimento di Fisica and INFN -- Sez. di Milano,
Universit\`a degli Studi di Milano, via Celoria 16, I-20133 Milano,
Italy\\
$^2$Max-Planck-Institut f\"ur Metallforschung,
Heisenbergstr. 3, D-70569 Stuttgart, Germany and
Institut f\"ur Theoretische und Angewandte Physik,
Universit\"at Stuttgart,
Pfaffenwaldring 57, D-70569 Stuttgart, Germany\\
$^3$Dipartimento di Matematica Applicata and INFN -- Sez.
di Pisa, Universit\`a degli Studi di Pisa, I-56100 Pisa, Italy \\
$^4$Dipartimento di Fisica and INFN -- Sez. di Roma I, Universit\`a degli Studi
di Roma ``La Sapienza'', I-00185 Roma, Italy
}
\begin{abstract}
We investigate the critical behavior of the two-dimensional 
randomly driven lattice gas, in which particles 
are driven along one of the lattice axes by an infinite external field with
randomly changing sign. 
A finite-size scaling (FSS) analysis 
provides novel evidences that this model 
{\it is not} in the same universality class as the driven lattice gas with a
constant drive (DLG), contrarily to what has been recently reported in
the literature.
Indeed, the FSS functions of {\it transverse observables}
(i.e., related to order-parameter fluctuations with wave
vector perpendicular to the direction of the field) 
differ from the mean-field behavior predicted and 
observed within the DLG universality class.
At variance with the DLG case, FSS is attained on 
lattices with fixed aspect ratio and anisotropy exponent equal to $1$ 
and the transverse Binder cumulant does not vanish at the critical
point.
\end{abstract}
%
\maketitle
  Phase transitions are characterized by a drastic change in the
  macroscopic behavior of many-body interacting systems when control
  parameters are varied. 
  In the case of critical phenomena the onset of the ordered phase is
  accompanied by fluctuations on all length scales.
  In spite of the difficulties in accounting efficiently for such coupled 
  fluctuations, we have currently a deep and detailed
  understanding of the collective behavior of a large class of systems 
  {\it in thermal equilibrium} thanks to a variety of results and methods, both
  analytical and numerical.
  Critical collective behaviors, 
  on the other hand, are also observed in the steady
  states of systems {\it far from thermal equilibrium}~\cite{Zia}.
  In contrast to equilibrium cases, 
  the stationary probability distribution is not known {\it a priori}
  and the possible occurrence and nature of a phase transition can no longer be
  determined by usual entropy-energy arguments.
  The absence of detailed balance, the generic algebraic decay
  of space-dependent correlations as functions of the distance, their
  strongly anisotropic scaling properties, 
  and the flux of particles or energy through the system are among
  the general features which make these problems particularly difficult and
  rich in phenomenology.
  Because of the lack of a general framework, it is still worth
  focusing on specific toy models in order to gain insight 
  which might possibly lead to a more comprehensive theory.
  Among the models characterized by nonequilibrium steady
  states the simplest is the uniformly driven lattice
  gas~\cite{KLS} (DLG), a generalization of the 
  Ising model to
  nonequilibrium conditions due to the action of an external
  nonconservative force, inducing a particle current through
  the system.
  Although the DLG was introduced more than 20 years ago, there is
  still room for debate on the nature and properties (in particular the
  {\it universality class}) of the nonequilibrium 
  critical behavior 
  observed upon changing the temperature. 
  At first, the relevant feature of the model was
  assumed  to be the presence of a particle current~\cite{Janssen86a}.
  However, more recently, this point has been criticized
  by arguing that the strong anisotropy is, instead, its qualifying
  character~\cite{Garrido,Achahbar}. In addition to the theoretical
  debate~\cite{dream-team}, seemingly contradictory evidences are also
  coming from Monte Carlo (MC)
  simulations~\cite{Achahbar,noi,Albano,reweight}.
  The Ising model can be driven to nonequilibrium conditions also by
  the coupling to two thermal baths
  at different temperatures~\cite{first}, 
  controlling the hopping rates of the 
  particles in different lattice directions. In a simpler version, the
  temperature affecting the jumps in one direction 
  is taken to be infinite.
  Hereafter we will refer to this direction as the {\it longitudinal}
  one ($\|$) whereas to the remaining as {\it transverse}
  ones ($\bot$).
  This model is equivalent to a DLG in which the
  microscopic external driving force is along the longitudinal
  direction, with infinite modulus and 
  a sign that is randomly
  chosen for each lattice site every time step (annealed randomness). 
  The resulting model is called the randomly driven lattice gas
  (RDLG)~\cite{RDLG}. At variance with the DLG, no net particle
  current is flowing through the system.
  MC simulations~\cite{two-reservoirs} indicate that in the RDLG
  {\it transverse fluctuations} of the order parameter 
  (i.e., the fluctuations with
  wave vector in the transverse direction) are not
  effectively described by a Gaussian model.
  Indeed, in two dimensions, the case we shall consider from now on, 
  MC simulations give $\pe{\nu} = 0.62(3)$ and  $\eta =
  0.13(4)$~\cite{two-reservoirs}. These estimates rely on a
  field-theoretical result for the {\it anisotropy exponent} 
  $\Delta \equiv (\pa{\nu}/\pe{\nu}) - 1$, i.e., 
  $\Delta = 1-\eta/2 \approx 1$~\cite{RDLG} which enters the
  finite-size scaling (FSS) Ans\"atze used to extract critical exponents.
  In Ref.~\cite{Achahbar} the numerical data for the RDLG and the DLG
  on the same finite lattices have been compared. According to these
  data the two models have the same finite-size scaling (FSS) behavior.
  If true, this implies that they belong to the same 
  universality class and thus the
  strong anisotropy and not the particle current is
  the relevant feature in determining the leading critical behavior of driven
  diffusive systems. The same conclusions has been drawn 
  in Ref.~\cite{Albano} by studying the short-time dynamics,
  although some points of the analysis therein remain unclear~\cite{comment}.
  Here we reconsider the problem presenting the results of 
  a new series of MC simulations of the RDLG. The critical
  behavior  of the system (proper of the thermodynamic limit) 
  is extracted from data on finite lattices by means of a FSS analysis that 
  does not require free parameters \cite{Caracciolo},
  in contrast with that previously employed~\cite{Leung92}. 
  We briefly recall the definition of the RDLG.
  Consider a rectangular lattice and for each site $x$ introduce an occupation 
  variable $n_{x}$, which can be either zero (empty site) 
  or one (occupied site). The external field 
  $E$ is acting along the longitudinal direction 
  but with random sign. The dynamics of the model is of Kawasaki type:
  A lattice link $\langle xy\rangle$ is randomly chosen, and, if
  $n_{x}\not= n_y$, a particle jump is proposed and then accepted with
  Metropolis probability $w(\beta \Delta H + \beta E \ell)$, where $\ell
  = (1,0,-1)$ for jumps (along, transverse, opposite) to $E$, $w(x) =
  \hbox{\rm min}\, (1,e^{-x})$, and $\Delta H$ is the variation 
  of the standard lattice-gas nearest-neighbor
  attractive interaction 
  $H = - 4\,\sum_{\langle xy\rangle} n_x n_y$ due to
  the proposed jump.
  The parameter $\beta$ plays the role of an inverse
  temperature. 
  In the DLG, $E$ is constant and time-independent. 
  Periodic boundary conditions in the direction of $E$ make it a
  nonconservative field and drive the system into a nonequilibrium stationary
  state. Although the boundary conditions are not so relevant in the
  RDLG, we will assume them periodic in all directions.

\begin{figure}
  \includegraphics[width=.4\textwidth]{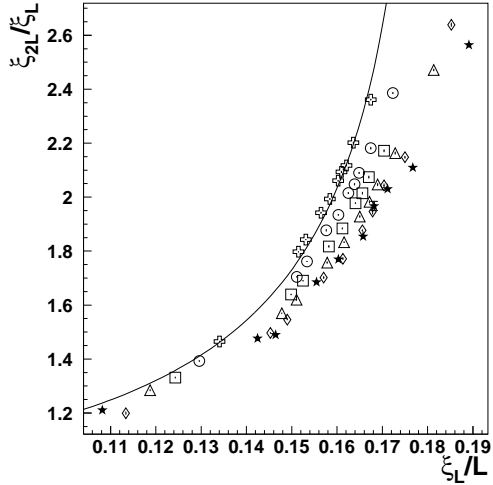}
  \caption{FSS plot of the transverse correlation length $\xi_L$ in the
  two-dimensional RDLG with fixed $S_{2} \approx
  0.200$. Crosses, circles, squares, triangles, diamonds, and stars
  correspond to lattices of increasing size $L = 14, 16, 18, 20, 22, 24$.
  The Gaussian behavior approached by the DLG is given by the solid line.
  }
  \label{fixedS2}
  \end{figure}

\begin{figure}
  \includegraphics[width=.4\textwidth]{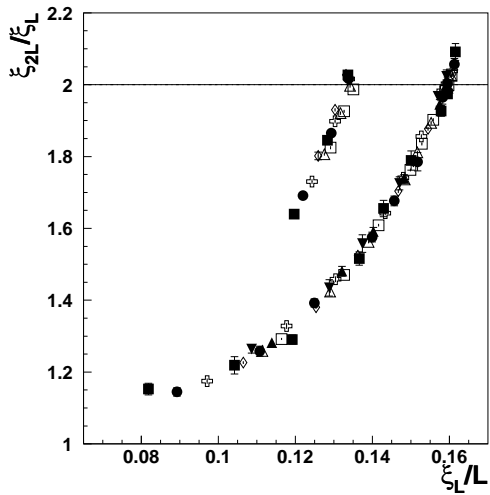}
  \caption{FSS plot of the transverse correlation length $\xi_L$ in the
  two-dimensional RDLG with fixed $S_1 = 0.223$ (upper set of points)
  and $S_1 = 0.326$ (lower set). Empty squares, empty triangles, diamonds, crosses, full circles, full squares, full upright triangles, and full downright triangles  correspond to lattices of increasing size $L = 20, 22, 24, 28, 32, 36, 40, 44$.}
  \label{fixedS1}
  \end{figure}

\begin{figure}
  \includegraphics[width=.4\textwidth]{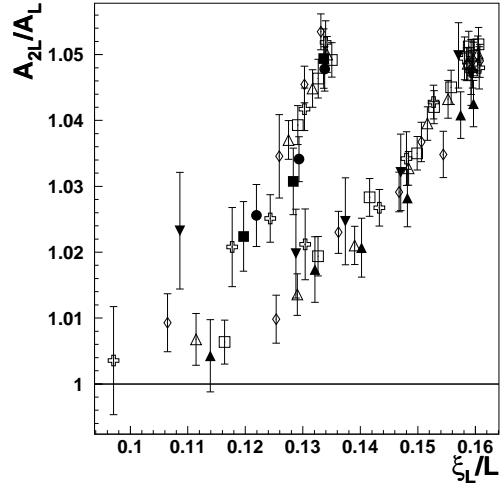}
  \caption{FSS plot of the ratio $\xi^{2}_L/\chi_L$ in the
  two-dimensional RDLG with fixed
  $S_{1} = 0.223$ (upper set of points) and $0.326$ (lower set). Symbols are as in Fig.~\ref{fixedS1}.}
  \label{A}
  \end{figure}

\begin{figure}
  \includegraphics[width=.4\textwidth]{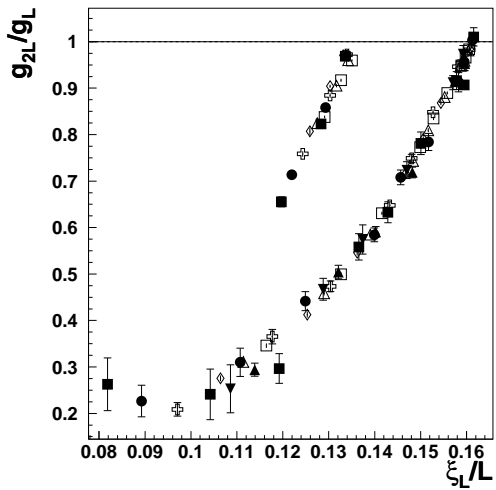}
  \caption{FSS plot of the transverse Binder cumulant $g_L$ in the
  two-dimensional RDLG with fixed
  $S_1=0.223$ (upper set of points) and $S_1=0.326$ (lower set). Symbols are as in Fig.~\ref{fixedS1}.}
  \label{g}
  \end{figure}

  At half filling, the RDLG undergoes a second-order phase transition. Indeed, 
  at high temperatures the steady state is disordered whereas at low 
  temperatures the system orders: The particles condense forming a strip 
  with interfaces parallel to $E$. 
  These two phases are separated by a phase transition 
  occurring at the critical value $\beta_c(E)$ depending on the field $E$. 
  Here we will concentrate on the particular case in which $E$ is infinite.
  For a strongly anisotropic system in $d$ dimensions, with finite size
  $\pa{L}\times\pe{L}^{d-1}$, the FSS limit corresponds to 
  $t\equiv 1-\beta/\beta_{c}\to 0$ (where $\beta_c$ is the bulk critical
  temperature), $\pa{L},\pe{L}\to\infty$, keeping fixed both
  combinations $t\pa{L}^{1/\pa{\nu}}$ and $t\pe{L}^{1/\pe{\nu}}$, and
  therefore also the so-called {\em aspect ratio} $S_{\Delta}=
  \pa{L}^{1/(1+\Delta)}/\pe{L}$~\cite{Binder}. 
  Accordingly, the FSS analysis of numerical data generally requires an
  {\it a priori} knowledge of the exponent $\Delta$~\cite{our}.
  It would be a real step towards a better understanding of
  nonequilibrium phase transitions to have at disposal FSS in a form
  suitable for these systems, reliable and powerful enough to disentangle
  those key features which might be buried in tiny differences when the
  volume of the samples are increased.
  In this direction we have already performed a detailed study of the
  FSS of the DLG~\cite{noi} by using the general strategy
  introduced in Ref.~\cite{Caracciolo},
  confirming the mean-field behavior of transverse
  fluctuations, with $\Delta=2$, in agreement with the predictions 
  of Ref.~\cite{Janssen86a}. 
  It is therefore a crucial test to examine by the same method also the
  RDLG.
  For previous studies of FSS in strongly anisotropic systems
  see Refs.~\cite{Binder,Leung92}.
  We will show that the method is sensible enough to highlight the
  differences in the critical behavior of the DLG and RDLG (contrarily to the
  claims in Ref.~\cite{Albano-comment}), to an extent
  that goes beyond the numerical differences in the critical exponents
  and probes the spatial structure of correlations.
  In spite of the generic power-law decay 
  of the two-point correlation function $\langle n_x n_0\rangle$ for large
  $x$~\cite{two-reservoirs},
  it is possible to define a finite-volume correlation length \cite{noi}.
  Given the Fourier transform $G(q)$ of $\langle n_x n_0\rangle$, 
  one considers 
  $\pe{G}(q) \equiv G(\left\{\pa{q}=0,\pe{q}=q\right\})$ 
  and defines the correlation length
  \begin{equation} 
  \label{defxi} 
  \xi_L \equiv 
  \sqrt{\frac{1}{\hat{q}_{3}^{2} - \hat{q}_{1}^{2}} 
  \left[ \frac{\pe{G}(q_{1})}{ \pe{G}(q_{3})} -1\right] },
  \end{equation} 
  where $\hat{q}_{n}=2\sin{q_n/2}$ is the lattice momentum 
  and $q_n = 2\pi n/L_\bot$. 
  Hereafter we will denote $L_\bot$ simply by $L$.
  In Fig.~\ref{fixedS2} we report the FSS plot of
  the ratio $\xi_{2L}/\xi_L$, where $\xi_{2L}$ and $\xi_L$
  are computed at the same temperature but lattice sizes $2 L$ and
  $L$, respectively, keeping constant the aspect ratio with 
  $\Delta = 2$, i.e., $S_2 \simeq 0.200$~\cite{correlation-length}. 
  For comparison we report as a solid line the mean-field prediction
  which is approached by the DLG data on the larger lattices~\cite{noi}.
  In the present case, deviations from the mean-field behavior increase with
  increasing lattice sizes. 
  Note that, if $S_1$ is the correct aspect ratio for the model,
  then one observes the crossover 
  towards the FSS of the model in the strip geometry $L_\bot =
  \infty$, when keeping $S_2$ constant and $\xi_L/L$ fixed~\cite{our}. 
  Accordingly, the points corresponding to larger lattices in
  Fig.~\ref{fixedS2} eventually accumulate on some limiting curve as $L$
  increases, in agreement with the predictions~\cite{noi_futuro}
  based on the field-theoretical model of Ref.~\cite{RDLG}.
  Fig.~\ref{fixedS1} refers to geometries with $\Delta = 1$,
  i.e., $S_{1}\approx 0.223$ (upper set of points) and  $S_{1}\approx
  0.326$ (lower set). Note that we used $\Delta=1$, although 
  a correction of the order of $\eta/2$ to this value is expected.
  In these cases we have been able to reach $L=88$.
  In contrast with the mean-field behavior, which does not depend on 
  the specific value of $S_\Delta$, 
  now we do expect a dependence of the FSS
  curves on the actual value of $S_1$. 
  The critical properties can be extracted from the
  previous plot as follows. For a given $S_1$, 
  consider the scaling function 
  $\xi_{2L}/\xi_{L} = F(z)$ as a functions of $z=\xi_L/L$ and
  expand it around $z^*$, which is 
  defined as the point such that $F(z^*) = 2$, 
  i.e., as
  the value of $\xi_L/L$ at the critical temperature. Denoting $\delta z = z
  -z^*$, one finds~\cite{noi}
  \begin{equation}
  \begin{split}
  F(z) & = F(z^{*}) + F'(z^{*}) \delta z+ O[(\delta z)^2]\\
  &  = 2 + \frac{2}{z^{*}} \left(2^{1/\pe{\nu}}- 1\right) \delta z+
  O[(\delta z)^2] \;.
  \end{split}
  \end{equation}
  A linear fit of our data gives
  $z^* = 0.1337(3)$ for $S_1 = 0.223$ and $z^* = 0.1594(1)$ for $S_1 =
  0.326$, with the same $\pe{\nu} =  0.61(3)$. The corresponding 
  critical temperature is the same in the two cases.
  Note that $z^{*}$ for $S_{1}= 0.326$ is almost equal to the mean-field value
  $1/(2 \pi)$~\cite{noi}. Indeed, we had chosen this value of
  $S_{1}$ in order to be very close to the mean-field predictions and
  test whether the FSS method employed is able to detect the differences.
  These results suggest that at variance with the DLG, where mean-field
  scaling at fixed $S_2$ is observed, in the
  present case scaling is attained only at fixed $S_1$ and is not
  compatible with mean-field behavior.
  Indeed, not only does $z^*$ depend on the geometry, 
  but also the critical exponent $\pe{\nu}$ differs from the Gaussian value 
  1/2.
  The qualitative
  dependence of $z^*$ on $S_1$ is accounted for~\cite{noi_futuro} by the
  field-theoretical model of Ref.~\cite{RDLG}.
  A similar analysis, with similar results, has been performed for the
  susceptibility $\chi_L \equiv \pe{G}(q_1)$~\cite{noi_futuro}.
  We find more instructive to present data for the ratio $A_L \equiv
  \xi_L^2/\chi_L$, which is independent of $\xi_L/L$ (for $L$ large
  enough) whenever the critical exponent $\eta$ vanishes. 
  In Fig.~\ref{A} we present the FSS data of this observable, for the
  two values of $S_{1}$ previously considered. 
  At variance with the DLG, where we got $A_{2L}/A_L \simeq 1$, 
  here we see a pronounced and systematic dependence on $\xi_L/L$ and on
  the actual value of $S_1$. 
  For the two different values of $z^{*}(S_1)$ we do find the same value
  for $A_{2L}/A_L \simeq 1.05$,
  which is equal to $2^\eta$~\cite{noi} and leads to the estimate
  $\eta = 0.07(1)$.
  Further evidence of differences in the critical behavior of the
  DLG and of the RDLG is provided by the transverse Binder cumulant 
  $ g_{L}   \equiv - G^{(4)}_\bot(q_1, q_1, -q_1, -q_1)/[N_L\,G^2_\bot(q_1)]$,
  where $G^{(4)}_{\bot}$ is the Fourier transform of the four-point
  connected correlation function at a given $\beta$ and lattice geometry,
  computed at the first allowed transverse momentum $q_1$, and
  $N_L$ is the total number of spins in the lattice. 
  The FSS plot in Fig.~\ref{g} 
  shows clearly that for both the values of $S_1$ considered, 
  $g_{2L}/g_{L} \simeq 1$ at the corresponding $z^*(S_1)$. 
  Therefore, $g_L$ at the critical point does not vanish in the
  thermodynamic limit, at variance with what has been found for the
  DLG~\cite{noi}.
  The estimated values of the critical exponents are in good agreement with
  previous numerical findings (although our result for $\eta$ 
  is smaller than that reported in Ref.~\cite{two-reservoirs}) and
  theoretical estimates based on the field-theoretical model of
  Ref.~\cite{RDLG}. Moreover, the {\it universal} FSS
  functions for the correlation length agree with those computed in 
  field theory at first order in an $\epsilon$ expansion around the
  upper critical dimension $d=3$~\cite{noi_futuro}.
  In conclusion, we have shown that the FSS approach as devised in
  Ref.~\cite{Caracciolo} is sensible enough to distinguish clearly
  between the critical behavior of the DLG and of the RDLG, two systems
  which in particular geometries may exhibit quite similar
  behavior for relatively small volumes and not too close to the critical
  temperature.
  Therefore, on one side we have a sound numerical method to examine
  also nonequilibrium critical phenomena, on the other we have
  eventually established that the key features of these two models are
  different, as
  they do not belong to the same universality class. The agreement with
  the field-theoretical analysis of
  Refs.~\cite{Janssen86a,RDLG,two-reservoirs} suggests that indeed the
  leading critical behavior of the DLG is governed by
  the presence of a particle 
  current whereas that one of the RDLG is dominated by strong anisotropy.
  %
  %

  \end{document}